\documentclass[11pt]{article}
\usepackage{graphicx,amsmath,bm, amsthm,mathrsfs,amssymb, braket, verbatim}
\usepackage{caption}
\usepackage{subcaption}
\usepackage[usenames]{color}
\usepackage{ulem,mathtools}
\usepackage{pdfpages}
\usepackage{lscape}
\usepackage{cite}
\usepackage{authblk}
\usepackage{cite}
\usepackage{tcolorbox}
\usepackage[section]{placeins}
\usepackage{placeins}

\newcommand{\ee}{\end{equation}}

\newcommand{\reff}[1]{(\ref{#1})}
\newcommand{\beq}{\begin{equation}}
\newcommand{\eeq}[1]{\label{#1}\end{equation}}
\newcommand{\beqa}{\begin{eqnarray}}
\newcommand{\eea}{\end{eqnarray}}
\newcommand{\eeqa}[1]{\label{#1}\end{eqnarray}}
\newcommand{\beg}{\begin{equation*}}
\newcommand{\eeg}{\end{equation*}}

\newcommand{\bsplit}{\begin{split}}
\newcommand{\esplit}{\end{split}}

\usepackage{circuitikz} 
\usepackage[capposition=bottom]{floatrow} 
\usepackage{epigraph} 
\usepackage{appendix}
\usepackage{rotating}
\allowdisplaybreaks

\title{Transition of a superposition of states under a delta-function pulse in a two-level system}
\author[]{Ariel Edery\thanks{aedery@ubishops.ca}}
\affil[]{Department of Physics and Astronomy, Bishop's University, 2600 College Street, Sherbrooke, Qu\'{e}bec, Canada, J1M 1Z7.\vspace{1em}}
\begin{document}
\date{}
\maketitle
\begin{abstract}
Under a time-dependent perturbation, it is common to calculate the probability of a transition from one eigenstate to another eigenstate of a quantum system. Here we study the transition from a \textit{linear superposition of eigenstates} to an eigenstate under a delta-function pulse.  We consider a two-level system with energy levels $E_1$ and $E_2$ and obtain exact analytical expressions for the coefficients $c_1$ and $c_2$ of the final state. The expressions are general since the coefficients $\alpha_1$ and $\alpha_2$ of the initial superposition state are free parameters constrained only by $|\alpha_1|^2+ |\alpha_2|^2=1$.  This opens up new possibilities and in particular allows for an abrupt transition to a definite eigenstate with unit probability. We obtain a general analytical expression for the probability $P_{\alpha_1,\alpha_2 \to 2}$  of an initial superposition state to transition to the second eigenstate. Armed with this general expression we study some interesting special cases. With a delta-function pulse, the transitions are abrupt/instantaneous and we show that they do not depend on the energy gap $E_2-E_1$ and hence on the relative phase between the two eigenstates. For specific values of the interaction strength $\beta$, the initial superposition transitions abruptly to a definite eigenstate with unit probability. We discuss the  similarities and differences such a transition has with the collapse of the wavefunction familiar in the context of a measurement.   

\end{abstract}
\setcounter{page}{1}
\newpage
\section{Introduction}\label{Intro}

Under a time-dependent interaction $H'(t)$ a quantity that has been of interest is the transition probability in going from one eigenstate to a second eigenstate of a time-independent Hamiltonian $H^0$. In such a case, the particle starts off in an initial state which is an eigenstate which could, for example, be the ground state of an atom, molecule, etc. One is however free to consider transitions where the initial state is a \textit{linear superposition of eigenstates}. This is more general and allows one to investigate physical transitions that would otherwise not be possible.        

In this work, we study the effects of a time-dependent interaction on a two-level system consisting of two energy eigenstates of a time-independent Hamiltonian $H^0$. We consider transitions where the initial state is a linear superposition of the two eigenstates. Moreover, the time-dependent interaction used is a delta-function pulse. This causes an abrupt/instantaneous transition from the initial superposition to a final state of the two-level system. In particular, for certain strengths of the delta-function pulse, the initial superposition can transition abruptly to a definite eigenstate i.e. with unit probability. The delta-function pulse is treated as a sequence of functions $q_n(t)$, where the pulse width gets smaller and its height gets taller as $n$ increases while the area under the pulse, its integral over time, stays constant. In the limit as $n$ tends to infinity, $q_n(t)$ approaches a Dirac delta function. The $q_n(t)$ are actually defined via their integrals and in calculations $q_n(t)$ will typically appear in the integrand of an integral. One does not encounter any infinities in calculations because one does not take the infinite $n$ limit of $q_n(t)$ itself but of the integral. The integral yields a finite function of $n$ and taking the infinite $n$ limit of that function yields a finite result. Though our motivation for using a delta-function pulse is to cause an abrupt/instantaneous transition, it has recently been pointed out \cite{Granot} that it also leads to one of the few quantum dynamical systems that can be solved exactly, analytically. 

The two-level system we consider has energies $E_1$ and $E_2$ (one can assume $E_2>E_1$). The coefficients appearing in the linear superposition associated with the initial state are $\alpha_1$ and $\alpha_2$ where $|\alpha_1|^2$ and $|\alpha_2|^2$ are the probabilities of measuring $E_1$ and $E_2$ respectively before the onset of the interaction. The time-dependent interaction $\hat{H}'$ has off-diagonal elements $H'_{12}=H'_{{21}^*} =\beta\,q_n(t)$ where $\beta$ is the interaction strength and $q_n(t)$ approaches a delta-function pulse at $t=0$ in the large (infinite) $n$ limit. For our purpose, the diagonal elements of $\hat{H}'$ are not required and for simplicity we set them to zero: $H'_{11}=H'_{{22}} =0$. One goal is to solve for the two coefficients appearing in the final state after the delta-function pulse acts at $t=0$ ; these are labeled $c_1(t>0)$ and $c_2(t>0)$. The coupled set of first order equations that mix $c_1(t)$ and $c_2(t)$ can be turned into second order equations for either $c_1(t)$ or $c_2(t)$ separately. We will see that this can be solved exactly in the large (infinite) $n$ limit i.e. for the delta-function pulse. For concreteness, our calculations are performed using Gaussians for the sequence of functions $q_n(t)$. We obtain \textit{exact analytical expressions} for $c_1(t>0)$ and $c_2(t>0)$ as a function of the interaction strength $\beta$ and the initial coefficients $\alpha_1$ and $\alpha_2$. In appendix A, we solve the coupled set of first order equations using a different method that does not involve combining them into a second order equation. The expressions obtained for $c_1(t>0)$ and $c_2(t>0)$ using this method match our original expressions providing a strong confirmation of our exact analytical results. 

An important physical consequence of using the delta-function pulse is that the coefficients $c_1(t>0)$ and $c_2(t>0)$ do not depend on the energy gap $E_2-E_1$ between the two states. The coupled equations for the coefficients contain the relative phase between the two eigenstates, a term $e^{i\,\omega_0\,t}$ where $\omega_0=(E_2-E_1)/\hbar$. In the limit as $n$ tends to infinity, which yields a delta-function pulse at $t=0$, the coefficients have no dependence on $\omega_0$ and hence on the energy gap. The relative phase is lost in an abrupt transition. Interestingly, the interaction with the environment in the decoherence framework leads to a very rapid loss in the relative phase relationship of the original pure state. This is somewhat intriguing and discussed further in the conclusion.     
 
Armed with the exact expressions for $c_1(t>0)$ and $c_2(t>0)$ we calculate the transition probability $P_{\alpha_1,\alpha_2 \to 2}$ in going from the initial superposition of states to the second eigenstate under the delta-function pulse. The expression for the probability is \textit{general} in the sense that they are functions of the coefficients $\alpha_1$ and $\alpha_2$ associated with the initial state. These are free parameters constrained only by $|\alpha_1|^2 + |\alpha_2|^2=1$. Our general expression is used to study some interesting special cases. It is also used to evaluate cases previously studied in order to compare results. A quantity of particular interest that has been studied in the past is the probability $P_{1 \to 2}$ for a transition from the first eigenstate to the second eigenstate (see \cite{Quantum}). We can obtain this by setting $\alpha_2=0$ and $|\alpha_1|=1$ in our general expression. We find that our results are in agreement with previous work in the limit as the pulse width tends to zero i.e. limit of a delta-function pulse. In previous work where a pulse of finite width was considered, the energy gap appears in the expression $P_{1 \to 2}$. However, we show that it disappears in the limit as the pulse width tends to zero. This confirms in an explicit and independent fashion our finding that the coefficients $c_1(t>0)$ and $c_2(t>0)$ do not depend on the energy gap $E_2-E_1$ for a delta-function pulse.          

The general expressions for the probabilities allow us to explore a special case where the initial superposition transitions abruptly under the delta-function pulse to a definite eigenstate with unit probability i.e. $|c_1(t>0)|^2=1$ assuming the definite eigenstate is the one with energy $E_1$. This occurs for specific values of the magnitude of the interaction strength $|\beta|$ and depends on $|\alpha_1|$. We provide a plot of $|\beta|$ as a function of $|\alpha_1|$. The magnitude of the slope increases with $|\alpha_1|$ and we discuss the implications of this.  

When would we expect a delta-function pulse to be of practical use? In the lab, an ultra-short laser pulse (pulse duration of femtosecond or attosecond) is often used for inducing atomic and molecular transitions as well as for causing physical transformations in solids \cite{Gogyan, Gamaly,Makarov}. Though these pulses are of finite duration, they are so short that all the major relaxation times associated with the physical system are larger by a few orders of magnitude \cite{Gamaly}. Practically speaking, this implies the ultra-short laser pulse acts almost instantaneously. In such cases, a delta-function pulse can be useful as an approximation to the real pulse, as this can simplify the theoretical analysis. Moreover, there is a physical process that is assumed in quantum mechanics to occur \textit{instantaneously}: this is the collapse of the wavefunction where a linear superposition of states is projected instantaneously (collapses) to a single eigenstate after a measurement. Under a delta-function pulse, the transition is also instantaneous. In section 4 we discuss how the abrupt/instantaneous transition from a linear superposition of states to a definite eigenstate with unit probability under a delta-function pulse has some similarities and some differences with the collapse of the wavefunction induced by a measurement.  
  
Our initial state is a superposition. Experimentally, this is not confined to the microscopic realm but can be realized macroscopically. There has been significant recent advances in the controlled preparation of macroscopic quantum superpositions in the lab. In particular, macroscopic superpositions via quantum tunneling have been realized  
\cite {Zhang}. In that work, high-mass spatial entanglement via the quantum tunneling of ultra-cold atoms in optical lattices was achieved. In \cite{Haine}, atoms were made to tunnel together as a bound cluster and this enabled the generation of massive cat states: the quantum superposition of two macroscopically distinct states. A two-level macroscopic quantum state was also achieved using a superconducting quantum interference device (SQUID) \cite{Friedman} ; this was placed in a quantum superposition of two opposing magnetic-flux states. These experiments show that a controlled initial superposition of states is not only something which is experimentally feasible but that, remarkably, it can even be produced on a macroscopic scale. At the microscopic level there has also been recent interest in studying high-energy quantum superpositions \cite{Canas} where the total probability density for a superposition of a large number of states converges exactly to the uniform classical distribution.       

The effect of a delta-function pulse in a two-level system was recently investigated in \cite{Granot}. The system studied consisted of a quantum dynamical model where the Hamiltonian includes diagonal elements with a delta-function pulse $V(t)$ and off-diagonal elements with constant coupling. The author obtains an exact analytical solution something which is rare for systems where the energy gap follows a pulse-like profile (for analytical results in other two-level model systems see 
\cite{FT,Carrol,Jakub,Bambini1,Bambini2,Garraway}). The analytical results in \cite{Granot} agree strongly with those obtained numerically for short finite duration pulses and reveal criteria for suppressed transmission (i.e. criteria for no transition so that it remains in the initial state) even though the delta-function pulse does not satisfy the slowly-varying constraint on which such criteria were previously found \cite{Granot2}. The exact solution also provides a way to analyze suppressed transmission in more complex systems like those that are harmonically driven (e.g. \cite{Grossmann1, Grossmann2}). The quantum dynamical model in \cite{Granot} differs from ours in a few regards and the focus of the study is different. First, the energy gap varies with the pulse whereas in our case it is constant. Secondly, the coupling is constant whereas in our case it is time-dependent i.e. the delta-function pulse appears in the off-diagonal elements (i.e. as part of the coupling). These differences lead to significantly different analytical expressions for the two coefficients $c_1(t)$ and $c_2(t)$ of the final state. The transition probabilities studied in \cite{Granot} are between two quasi-states $\left|1\right\rangle$ and $\left|2\right\rangle$ whereas we focus on transition probabilities between a superposition of two eigenstates and an eigenstate. In particular, there exist values of the interaction where the initial superposition transitions to a definite eigenstate with unit probability. Other work involving transitions in a two-level system have concentrated on transition probabilities between two eigenstates (or quasi-states) and used finite pulses instead of delta-function pulses (see \cite{Quantum} for a review). In our work, a delta-function pulse is a physical requirement and not optional as it is the interaction that causes an abrupt/instantaneous transition. 

Our paper is organized as follows. In section 2 we derive the well-known coupled set of first-order equations for the coefficients under a time-dependent perturbation. This section is naturally brief. In section 3 we solve the coupled set of equations for a delta-function pulse acting at $t=0$ and obtain general analytical expressions for the coefficients $c_1(t>0)$ and $c_2(t>0)$. In appendix A we derive the same expressions using a different method. In section 3.1 we obtain a general formula for the transition probability $P_{\alpha_1,\alpha_2 \to 2}$ and apply it to special cases of interest. Section 3.2 is devoted to the energy gap independence of the probability under a delta-function pulse. Section 4 discusses the scenario where the initial superposition transitions abruptly to a definite eigenstate with unit probability. Similarities and differences with a real measurement are discussed. Section 5 is the conclusion where we summarize and discuss our results as well as decoherence.        

\section{Two-level system under a time-dependent perturbation}\label{S2}
We begin by deriving a well-known exact expression in time-dependent perturbation theory; this will not only help establish the notation but will allow us to explain more clearly the new results we obtain. The derivation in this section will naturally be brief.  We consider a two-level system where $\psi_1$ and $\psi_2$ are eigenstates of the time-independent Hamiltonian $\hat{H}^0$:
\beq
\hat{H}^0\,\psi_1 =E_1\, \psi_1 \quad;\quad \hat{H}^0\,\psi_2 =E_2\, \psi_2 \,.
\eeq{Two}
$\psi_1$ and $\psi_2$ are orthonormal
\beq
\left\langle \psi_a |\psi_b \right \rangle= \delta_{ab}
\eeq{Ortho}
where $a$ and $b$ can take on the values of $1$ or $2$. The wavefunction $\Psi(t)$ (space dependence may exist but is not stated explicitly) can be written as usual as a linear combination of the two states with their time-dependent phase factor:
\beq 
\Psi(t) = c_1\, \psi_1\, e^{-i E_1\,t/\hbar} +c_2 \,\psi_2 \,e^{-i E_2\, t/\hbar}\,.
\eeq{Psit}
Normalization implies that $|c_1|^2+|c_2|^2$=1. Let $\hat{H}'(t)$ be a time-dependent perturbation so that the full Hamiltonian is 
\beq
\hat{H}=\hat{H}^0 + \hat{H}'(t)\,.
\eeq{FullH} 
Since $\psi_1$ and $\psi_2$ form a complete set we can still express $\Psi(t)$ as a linear combination of them except that now the coefficients $c_1$ and $c_2$ depend on time \cite{Weinberg}:
\beq 
\Psi(t) = c_1(t)\, \psi_1\, e^{-i E_1\,t/\hbar} + c_2(t) \,\psi_2 \,e^{-i E_2\, t/\hbar}\,.
\eeq{Psit2}
We prefer not to absorb the phase factors into $c_1(t)$ and $c_2(t)$ since the phase factors would be present without a time-dependent perturbation. The goal is to now solve for $c_1(t)$ and $c_2(t)$ by applying Schr\"odinger's equation:
\beq 
i \hbar \frac{\partial \Psi}{\partial t} = \hat{H} \,\Psi \,.
\eeq{Psit3}
Substituting \reff{FullH} and \reff{Psit2} into \reff{Psit3} and using \reff{Two} yields the equation 
\begin{align}
i\,\hbar\,( \dot{c_1}\,\psi_1\, e^{-i E_1\,t/\hbar} + \dot{c_2}\,\psi_2\, e^{-i E_2\,t/\hbar}) = c_1 \,\hat{H}'\, \psi_1\,e^{-i E_1\,t/\hbar} + c_2 \,\hat{H}'\,\psi_2\,e^{-i E_2\,t/\hbar} \,.
\label{Quad}
\end{align}  
It is convenient to define
\beq
H'_{ab}=\left\langle \psi_a |\hat{H}'|\psi_b \right \rangle
\eeq{Hab}
where $H'_{ba}=(H'_{ab})^{*}$ since $\hat{H}'$ is a hermitian operator. 
By taking the inner product with $\psi_1$ and with $\psi_2$ separately in \reff{Quad} and using the orthonormality condition \reff{Ortho} we obtain the following equations:
\beq
\dot{c_1}=\frac{-i}{\hbar}\Big(c_1\, H'_{11} +c_2 \,H'_{12}\, e^{-i (E_2-E_1)\,t/\hbar}\Big)
\eeq{c1dot}
\beq 
\dot{c_2}=\frac{-i}{\hbar}\Big(c_1\, H'_{22} +c_1 \,H'_{21}\, e^{i(E_2-E_1)\,t/\hbar}\Big)\,.
\eeq{c2dot}  
Equations \reff{c1dot} and \reff{c2dot} are \textit{exact}\cite{Weinberg}; no approximations have been made. We will be using these exact equations in the next section.

\section{Expressions for $c_1(t)$ and $c_2(t)$ after a delta-function pulse at $t=0$}\label{S3}
Consider the time-dependent perturbation 
\beq
\hat{H}'=\hat{V} \,q_n(t)
\eeq{HH}
where $q_n(t)$ is a sequence of functions labeled by the positive integer $n$ and depends on time only. The operator $\hat{V}$ is time-independent so that the time-dependence of $\hat{H}'$ stems from $q_n(t)$ only. The operator $\hat{V}$ acts on the time-independent states $\psi_1$ and $\psi_2$. The functions $q_n(t)$ are well-behaved functions that have a maximum at $t=0$. They have the following properties:
\begin{align}
&\int_{-\infty}^{\infty} q_n(t)\, dt= 1 \quad \forall \,n \in \mathbb{Z}^+ \label{Prop1}\\
&\lim_{n\to \infty} \int_{-\infty}^{\infty} q_n(t)\, f(t) \, dt= f(0) \,.\label{Prop2}
\end{align}
The first property \reff{Prop1} states that, regardless of the value of $n$, the ``area" under the curve is always unity. The second property \reff{Prop2} applies to any well-behaved function $f(t)$. Note that the infinite limit in \reff{Prop2} is taken after the integral is performed; the integral is a function of $n$ and finite. When the infinite $n$ limit is taken of the integral, it yields also a finite result. Therefore, no infinities are ever encountered. Since the limit yields the result $f(0)$ regardless of the function $f(t)$, this implies that the functions $q_n(t)$ become increasingly concentrated at $t=0$ as $n$ increases; they become thinner and higher near $t=0$ to maintain an area equal to one. The sequence of functions $q_n(t)$ with properties \reff{Prop1} and \reff{Prop2} can be viewed as representing a Dirac $\delta(t)$ at $t=0$ \cite{Arfken}. An example of a sequence of well-behaved functions $q_n(t)$ having properties \reff{Prop1} and \reff{Prop2} are the Gaussians       
\beq
q_n(t) =\frac{n}{\sqrt{\pi}}\, e^{-n^2 \,t^2}\quad \forall \,n \in \mathbb{Z}^+\,\,.
\eeq{func}
The above function is symmetric about $t=0$ and for a given positive integer $n$, has its maximum at $t=0$. It becomes increasingly concentrated at $t=0$ as $n$ increases; it becomes thinner and taller about $t=0$ while preserving its area.  The Gaussians \reff{func} act as a Dirac delta function at $t=0$ in the large $n$ limit.               
            
We are now ready to investigate the effects of the perturbation $\hat{H}'=\hat{V} \,q_n(t)$. From \reff{Hab} we have $H'_{ab}= V_{ab}\, q_n(t)$. Let  $V_{12}=V_{21}^* =\beta$ and $V_{11}=V_{22}=0$ ($V_{11}$ and $V_{22}$ are not necessary to create transitions between states so setting them to zero simplifies things). We will refer to $\beta$ as the interaction strength. It follows that $H'_{12}=\beta \,q_n(t)$, $H'_{21}=\beta^{*}\,q_n(t)$ and $H'_{11}=H'_{22}=0$. It is convenient to define the positive quantity $\omega_0 = (E_2-E_1)/\hbar$. Then equations \reff{c1dot} and \reff{c2dot} for $\hat{H}'=\hat{V} \,q_n(t)$ reduce to
\beq
\dot{c_1}=\frac{-i}{\hbar}\Big(c_2 \,\beta\, q_n(t)\, e^{-i\,\omega_0\, t}\Big)
\eeq{c1dot2}
\beq 
\dot{c_2}=\frac{-i}{\hbar}\Big(c_1 \,\beta^*\, q_n(t)\, e^{i\,\omega_0 \,t}\Big)\,.
\eeq{c2dot2}
The time-dependent interaction $\beta\, q_n(t)$ acts as a delta-function pulse at $t=0$ in the large (infinite) $n$ limit and causes an abrupt (basically instantaneous) transition from the initial state at $t<0$ to the final state at $t>0$. The coefficients for the initial state are constant in time and denoted $\alpha_1$ and $\alpha_2$:
\beq 
\Psi(t<0) = \alpha_1\, \psi_1\, e^{-i E_1\,t/\hbar} +\alpha_2 \,\psi_2 \,e^{-i E_2\, t/\hbar}\,.
\eeq{InitialP}
where $|\alpha_1|^2 +|\alpha_2|^2=1$. This means that in \reff{Psit2}
\beq
c_1(t<0)=\alpha_1 \text{ and } c_2(t<0)=\alpha_2\,.
\eeq{c2c1}
After the delta-function pulse acts at $t=0$, the coefficients in the final state are $c_1(t>0)$ and $c_2(t>0)$ and are also constant:
\beq 
\Psi(t>0) = c_1(t>0)\, \psi_1\, e^{-i E_1\,t/\hbar} + c_2(t>0) \,\psi_2 \,e^{-i E_2\, t/\hbar}
\eeq{FinalP}
The coefficients $c_1(t>0)$ and $c_2(t>0)$ in the final state will usually be different from the coefficients $c_1(t<0)=\alpha_1$ and $c_2(t<0)=\alpha_2$ of the initial state. However, in both cases the coefficients are constant, independent of time since the time-dependent interaction acts only at $t=0$ in the large (infinite) $n$ limit. 

Our goal is to solve the coupled set of first order equations \reff{c1dot2} and \reff{c2dot2} for $c_1(t>0)$ and $c_2(t>0)$ in terms of $\alpha_1$, $\alpha_2$ and the interaction strength $\beta$. We will use the Gaussians \reff{func} for $q_n(t)$. We begin by taking the derivative of $\reff{c1dot2}$ with respect to time. We can express $\dot{c_2}$ in terms of $c_1$ via \reff{c2dot2} and express $c_2$ in terms of 
$\dot{c_1}$ via \reff{c1dot2} to obtain the following second order equation for $c_1(t)$:
\beq
\frac{\ddot{c_1}}{n^2}+ \dot{c_1}\,\big(2\,t +\tfrac{i\,\omega_0}{n^2}\big) + \frac{1}{ \pi\,\hbar^2}\, c_1 \,|\beta|^2 \ e^{-2 \,n^2\,t^2}=0\,.
\eeq{c1Second}
Note that in the above equation the term $(2\,t +\tfrac{i\,\omega_0}{n^2})$  tends to $2\,t$ in the large (infinite) $n$ limit. We can therefore simply leave $2\,t$ in the brackets to solve the equations in the large $n$ limit. We could have anticipated this beforehand. In \reff{c1dot2} and \reff{c2dot2}, $q_n(t)$ will be concentrated near 
$t=0$ in the limit of large (infinite) $n$. It should be clear that in this limit we could have simply set $e^{\pm i\,\omega_0\, t}$ to unity since it is a smooth function with no singularities in the vicinity of $t=0$. In the large $n$ limit, \reff{c1Second} therefore reduces to    
\beq
\pi\,\hbar^2\,\ddot{c_1} + 2\,\pi\,\hbar^2\,\dot{c_1}\,n^2\,t + c_1 \,|\beta|^2\,n^2\, e^{-2 \,n^2\,t^2}=0\,.
\eeq{c1Second2}
The solution to the above differential equation is 
\beq
c_1(t)=b_1\cos \left(\frac{| \beta | \, \text{erf}\,(n\, t)}{2 \,\hbar }\right)+ b_2 \sin \left(\frac{| \beta |  \,\text{erf}\,(n\, t)}{2 \,\hbar }\right)
\eeq{c1t}
where $b_1$ and $b_2$ are integration constants and erf$\,(x)$ is the error function defined by
\beq
\text{erf}\,(x)=\frac{2}{\sqrt{\pi }}\int _0^x  e^{-t^2}\,dt\,.
\eeq{erf}
Substituting $c_1(t)$ given by \reff{c1t} into \reff{c1dot2} yields
\beq
c_2(t)= i\,b_2 \,\frac{| \beta |}{\beta} \cos \left(\frac{| \beta | \, \text{erf}\,(n \,t)}{2 \,\hbar }\right)-i\,b_1 \frac{| \beta |}{\beta}  \sin \left(\frac{| \beta |  \,\text{erf}\,(n\, t)}{2 \hbar }\right)\,.
\eeq{c2t}
The two initial conditions are that $\lim_{t\to -\infty}\,c_1(t)=\alpha_1$ and $\lim_{t \to -\infty}\,c_2(t)=\alpha_2$. In the large (infinite) $n$ limit where the interaction is a delta-function pulse, taking time back to $-\infty$ for the initial conditions is not required since there is no interaction until $t=0$. In other words, the initial state is the same in the entire region $t<0$ and the initial conditions can be written as $c_1(t<0)=\alpha_1$ and $c_2(t<0)=\alpha_2$. This can be seen mathematically from the fact that the following two limits yield the same answer:
\beq 
\lim_{t\to -\infty} \text{erf}\,(n\,t)=-1 \text{ and } \lim_{\substack{n \to \infty\\t<0}} \text{erf}\,(n\,t)=-1\,.  
\eeq{limiting}
Taking the above limit of $\text{erf}\,(n\,t)$ in \reff{c1t} and \reff{c2t}, the initial conditions yield the following two equations for the integration constants:
\begin{align}
b_1\cos \Big(\frac{| \beta |}{2 \,\hbar }\Big)- b_2 \sin \Big(\frac{| \beta |}{2 \,\hbar }\Big)&=\alpha_1 \\
i\,b_2 \,\frac{| \beta |}{\beta}  \cos \left(\frac{| \beta |}{2 \,\hbar }\right) +i\,b_1 \frac{| \beta |}{\beta}  \sin \left(\frac{| \beta |}{2 \hbar }\right)&=\alpha_2\label{c2init}\,.
\end{align}
Solving the above two equations yields
\begin{align}
b_1&=\alpha _1 \cos \Big(\frac{| \beta |}{2 \,\hbar }\Big)-i\, \alpha _2 \,\frac{\beta}{| \beta | }  \sin \Big(\frac{| \beta | }{2 \hbar }\Big)\label{b1A}\\
b_2&= -\alpha _1 \,\sin \Big(\frac{| \beta |}{2 \,\hbar }\Big)-i \,\alpha _2\, \frac{\beta}{| \beta | }\,\cos \Big(\frac{| \beta |}{2 \,\hbar }\Big)\,.\label{b2A}
\end{align}
Substituting $b_1$ and $b_2$ above into equations \reff{c1t} and \reff{c2t} and taking the large (infinite) $n$ limit with $t>0$, one obtains  
\begin{align}
c_1(t>0)&=\alpha _1 \cos \Big(\frac{| \beta | }{\hbar }\Big)-i \,\alpha_2 \,\frac{|\beta|}{\beta ^*}\, \sin \Big(\frac{| \beta | }{\hbar }\Big)\label{c1tg0}\\
c_2(t>0)&=\alpha _2 \cos \Big(\frac{| \beta | }{\hbar }\Big)-i\, \alpha_1 \,
\frac{|\beta|}{\beta } \, \sin \Big(\frac{| \beta | }{\hbar }\Big)\label{c2tg0}
\end{align}
where we used $\lim_{n \to \infty} \text{erf}\,(n\,t)=1$ for $t>0$. We have met our first goal of obtaining the coefficients of the final state in terms of the interaction strength $\beta$ and the coefficients $\alpha_1$ and $\alpha_2$ of the initial state. Note that $c_1(t>0)$ and $c_2(t>0)$ are constant, independent of time, as there is no longer a time-dependent interaction after the delta-function pulse at $t=0$. A non-trivial check on our final expressions \reff{c1tg0} and \reff{c2tg0} is that $|c_1(t>0)|^2 + |c_2(t>0)|^2$ should equal unity. In other words, the total probability should be conserved after the interaction. Summing the squares of the coefficients yields     
\beq
|c_1(t>0)|^2 + |c_2(t>0)|^2= (|\alpha_1|^2+|\alpha_2|^2)\Big[\,\cos^2 \Big(\frac{| \beta | }{\hbar }\Big)+ \sin^2 \Big(\frac{| \beta | }{\hbar }\Big)\,\Big]=|\alpha_1|^2+|\alpha_2|^2=1
\eeq{Norm}
where we used the fact that the initial state is normalized i.e. $|\alpha_1|^2+|\alpha_2|^2=1$. The sum of the square of the coefficients is indeed unity confirming that the total probability has been conserved after the interaction. In appendix A, using a different method,  we derive the expressions \reff{c1t0} and \reff{c2t0} for $c_1(t>0)$ and $c_2(t>0)$ respectively. These match the expressions \reff{c1tg0} and \reff{c2tg0} obtained above. This provides another strong confirmation of our exact expressions. 

\subsection{Transition probability: general expression}\label{Trans}

After the delta-function pulse at $t=0$ has acted on the system, a measurement of the energy (at $t>0$), yields either $E_1$ with probability $|c_1(t>0)|^2$ or $E_2$ with probability $|c_2(t>0)|^2$. Since $|c_1(t>0)|^2 =1- |c_2(t>0)|^2$ we can evaluate $|c_2(t>0)|^2$ only. This is the probability of a transition to the second energy eigenstate starting with the initial state \reff{InitialP} which is a linear superposition of the two eigenstates with coefficients $\alpha_1$ and $\alpha_2$. We can therefore label this transition probability as $P_{\alpha_1,\alpha_2 \to 2}$.  Multiplying \reff{c2tg0} by its complex conjugate we obtain
\begin{align}
P_{\alpha_1,\alpha_2 \to 2}&=|c_2(t>0)|^2\nonumber\\
&= |\alpha_2|^2 \,\cos^2 \Big(\frac{| \beta | }{\hbar }\Big) +  |\alpha_1|^2 \,\sin^2 \Big(\frac{| \beta | }{\hbar }\Big) + i\,|\beta|\,\sin \Big(\frac{2\,| \beta | }{\hbar }\Big) \Big(\frac{\alpha_1^*\,\alpha_2}{2\,\beta^*}-\frac{\alpha_2^*\,\alpha_1}{2\,\beta} \Big)\,.
\label{c2Sq}
\end{align}
The above result depends on the interaction strength $\beta$, the coefficients $\alpha_1$ and $\alpha_2$ but \textit{not} on the energy gap $E_2-E_1\,.$ The expression is \textit{general} in the sense that it is valid for any $\alpha_1$ and $\alpha_2$ (with constraint $|\alpha_1|^2+|\alpha_2|^2=1$).  

Let us now look at a case where the last term in \reff{c2Sq} is zero. This can occur, for example, when  $\sin \Big(\frac{2\,| \beta | }{\hbar }\Big)=0$ and hence when $|\beta | =\frac{\ell\,\pi\,\hbar}{2}$ where $\ell \in \mathbb{Z}^+$. Substituting $|\beta | =\frac{\ell\,\pi\,\hbar}{2}$ into \reff{c2Sq} we obtain
\begin{align}
P_{\alpha_1,\alpha_2 \to 2}&= |\alpha_2|^2 \,\cos^2 \Big(\frac{\ell\,\pi}{2 }\Big) +  |\alpha_1|^2 \,\sin^2 \Big(\frac{\ell\,\pi}{2 }\Big)\nonumber\\
&=
\begin{cases}
|\alpha_1|^2& \text{ if } \ell \in \mathbb{Z}^+ \text{ is odd}\\
|\alpha_2|^2 &\text{ if } \ell  \in \mathbb{Z}^+  \text{ is even}\,.
\end{cases}
\end{align}
This is quite a remarkable result. It states that if $|\beta |=\tfrac{\ell\,\pi\,\hbar}{2}$ with $\ell$ a positive even integer, then the probability is $|\alpha_2|^2$ which is the same as the probability of measuring $E_2$ in the initial state. \textit{It is as if no interaction occurred}. If we divide the positive even integers into $\ell=4\,n$
and $\ell=4\,n-2$ where $n$ is a positive integer then the initial state stays the same if $\ell=4\,n$ but changes sign if $\ell=4\,n-2$. Let us see this. If one substitutes the interaction $|\beta |=\tfrac{\ell\,\pi\,\hbar}{2}$ into the equations \reff{c1tg0} and \reff{c2tg0} for the coefficients $c_1$ and $c_2$ respectively, one obtains $(c_1,c_2)=(\alpha_1,\alpha_2)$ if $\ell=4\,n$ and $(c_1,c_2)= (-\alpha_1,-\alpha_2)$ if $\ell=4\,n-2\,.$ In the latter case, the interaction changes the state by flipping its sign but this does not affect the probabilities [however, it is worth noting that the sign change (a phase shift of $\pi$) can be detected \textit{relative to another state} via quantum interference]. If $\ell$ is odd something interesting happens: the initial roles of $\alpha_1$ and $\alpha_2$ are basically switched under the interaction. The probability of measuring $E_2$ becomes $|\alpha_1|^2$ and the probability of measuring $E_1$ becomes $|\alpha_2|^2$. A worthwhile question is whether such switching can find some application (practically, this would require an ultra-short pulse with a specific value of the strength).       
    
In the literature there have been studies of transitions in two-level systems with finite pulses (see \cite{Quantum} for a review). A quantity that has been calculated is the probability $P_{1 \to 2}$ of a transition from the first eigenstate with energy $E_1$ to the second eigenstate with energy $E_2$. The initial state in such a transition has $|\alpha_1|=1$ and $\alpha_2=0$. Substituting these values into \reff{c2Sq} yields 
\beq
P_{1 \to 2}=\sin^2 \big(\tfrac{| \beta | }{\hbar }\big)\,.
\eeq{P12A}
The above is our result for a delta-function pulse. The result for a finite pulse $\Omega(t)$ of width $T$ is (see \cite{Quantum} for details on the derivation):
\beq
P_{1 \to 2}=\frac{\sin^2 \big(\tfrac{1}{2}\,\pi \,\Omega_0\,T\big)}{\cosh^2\big(
\tfrac{1}{2}\,\pi \,\Delta_0\,T\big)}\,
\eeq{P12B}
where $\Delta_0=(E_2-E_1)/\hbar$ has dimension of frequency and is the constant energy gap $\Delta E=E_2-E_1$ in units of $\hbar$, $\pi\,\Omega_0\,T$ is the dimensionless (positive) area of the pulse $\Omega(t)$ (integral over time of the pulse) and $\Omega_0$ is the amplitude or height of the pulse with dimensions of frequency. By comparing the coupled first order equations, the real finite pulse $\Omega(t)$ corresponds to our quantity $2\,\beta\,q_n(t)/\hbar$ with $n$ assumed large. The integral over time yields $2\,|\beta|/\hbar$ which becomes the dimensionless area identified with $\pi\,\Omega_0\,T$ (we took $\beta$ here to be positive and real to obtain a positive real area. Hence $\beta$ was replaced by $|\beta|$ for purposes of comparison). The delta-function pulse is equivalent to taking the limit as the pulse width $T$ tends to zero while maintaining the area under the pulse. Therefore the area $\pi\,\Omega_0\,T$ remains constant under this limit (i.e. as $T$ gets smaller, $\Omega_0$ gets larger so that their product stays constant). Evaluating \reff{P12B} in the $T\to 0$ limit we obtain 
\beq
\lim_{T\to 0}\, P_{1 \to 2}=\lim_{T\to 0} \,\frac{\sin^2 \big(\tfrac{1}{2}\,\pi \,\Omega_0\,T\big)}{\cosh^2\big(\tfrac{1}{2}\,\pi \,\Delta_0\,T\big)}= \sin^2 \big(\tfrac{1}{2}\,\pi \,\Omega_0\,T\big)=\sin^2 \big(\tfrac{| \beta | }{\hbar }\big)
\eeq{T0}
where we used the fact that the area $\pi\,\Omega_0\,T$ had been identified with the area $\tfrac{2\,|\beta|}{\hbar}$. We also used that  $\lim_{T\to 0} \cosh^2\big(\tfrac{1}{2}\,\pi \,\Delta_0\,T\big)=1$. In \reff{T0} we recovered our result \reff{P12A} for the delta-function pulse as a limiting case of the finite pulse. This provides yet another confirmation of our general expressions \reff{c1tg0} and \reff{c2tg0}. 

\subsection{Energy gap independence of the transition probability under a delta-function pulse}

In a two-state system with energy eigenstates $E_1$ and $E_2$ (assume $E_2>E_1$), our physical intuition would tell us that the probability for a particle to transition from the eigenstate with energy $E_1$ to the eigenstate with energy $E_2$ under a pulse would decrease as the energy gap $\Delta E=E_2-E_1$ increases. Simply put, one would expect that as $E_2$ gets further away from $E_1$, then the probability of reaching it should decrease. The mathematics corroborates this intuition but only for a finite pulse: the transition probability $P_{1 \to 2}$ given by \reff{P12B} for a pulse with finite width $T$ decreases as the $\Delta_0$ increases (recall that $\Delta_0$ is the energy gap $\Delta E$ divided by $\hbar$).   

However, we saw in \reff{T0} that in the limit as the pulse width $T$ tends to zero 
(while maintaining a constant area $\pi\,\Omega_0\,T$) that the transition probability $P_{1 \to 2}$ has no dependence on $\Delta_0$ and hence no dependence on the energy gap.  A pulse with constant area but whose width tends to zero is precisely a delta-function pulse. So this limit on a more realistic finite pulse is a nice concrete way of seeing how the energy gap does not enter into the probability for a delta-function pulse. Moreover, the limit went beyond showing the energy gap independence. By taking the pulse width to zero in \reff{T0} we obtained the exact same probability as \reff{P12A} that was derived directly for a delta-function pulse. So there is a quantitative matching of the two probabilities. 

The energy gap independence is in accord with some of our previous analysis. The relative phase $e^{\pm \,i\, \omega_0\,t}$ where $\omega_0=(E_2-E_1)/\hbar$ appears together with the pulse interaction $\beta\,q_n(t)$ in the coupled set of equations \reff{c1dot2} and \reff{c2dot2} that govern the coefficients $c_1(t>0)$ and $c_2(t>0)$. The function $q_n(t)$ is given by the Gaussian \reff{func} and has a width about $t=0$ or standard deviation of $T= \frac{1}{\sqrt{2}\,n}$ . The relative phase \textit{needs time to evolve to have an effect on $c_1$ and $c_2$ and hence on the probabilities}. Such a change will occur if $n$ is finite so that the time $T$ that the pulse acts is finite. The energy gap should then appear in the probabilities as a function of $\omega_0\,T$. It is the quantity $\omega_0\,T$ not $\omega_0$ by itself that is of importance. That is why in \reff{P12B} the denominator comes with the product $\Delta_0\,T$ and not just $\Delta_0$ by itself. However, in the delta function limit where $n \to \infty$ and $T\to 0$, $q_n(t)$ is entirely concentrated at $t=0$ so that one can effectively replace the relative phase  $e^{\pm \,i\, \omega_0\,t}$ by unity in equations \reff{c1dot2} and \reff{c2dot2}. It is then clear that the energy gap will not appear in the probabilities for a delta-function pulse. This also means that the relative phase is in essence lost. These are important features of a delta-function pulse.   

There is a simple physically intuitive way to see that the energy gap should not enter into the probabilities for a delta-function pulse. The Fourier transform of the Dirac delta function $f(t)=\delta(t)$ is unity i.e.
\beq
\tilde{f}(\omega)=\int_{-\infty}^{\infty} \delta(t)\,e^{-i \,\omega\,t}\,dt=1\,.
\eeq{Four} 
This means that it contains all frequencies with equal strength; it ``sweeps" through all frequencies equally. Therefore, under a delta-function pulse, transitions will occur for \textit{all} energy gaps with \textit{equal} probability. Simply put, the probability is flat with respect to frequency and hence the energy gap.  This is of course the same thing as saying that it is independent of the energy gap. In equation \reff{P12A}, the transition probability $P_{1 \to 2}$ for a delta-function pulse with strength $\beta$ is given by $\sin^2 \big(\tfrac{| \beta | }{\hbar }\big)$. Though it depends on $| \beta |$, the probability is the \textit{same} regardless of the value of the energy gap $\Delta E=E_2-E_1$. A plot of the probability versus $\Delta E$ is a flat horizontal line. 

In the introduction we mentioned that the loss of the relative phase under a delta-function pulse brings to mind the rapid loss of the relative phase that occurs during decoherence when the system is coupled to the environment. As we stated in the introduction, decoherence will be discussed further in the conclusion.    
 
\section{Transition to a definite eigenstate with unit probability}\label{scenario}

In this section, we will see that the delta-function pulse can create an abrupt/instantaneous transition from a linear superposition of eigenstates to a definite eigenstate with unit probability. This case is interesting as it has two aspects in common with the collapse of the wavefunction induced by a measurement. First, in both cases the initial state is a linear superposition of eigenstates and the final state is a definite eigenstate. Secondly, in both cases the transition is abrupt or instantaneous. There are also two significant differences between the two cases. First, the abrupt transition to a definite eigenstate under a time-dependent potential in the context of Schr\"odinger's equation is a unitary reversible process whereas the collapse of the wavefunction induced by a measurement is a non-unitary irreversible process. We will show explicitly in section 4.1 that the transition to the definite eigenstate caused by the delta-function pulse is a reversible process. Secondly, and probably more important, under the evolution of Schr\"odinger's equation one can predict what the definite eigenstate will be whereas before a measurement, one only knows the probability of ending up in a particular eigenstate via the Born rule. So there are similarities and differences.    

The initial state of the two-level system is a \textit{linear superposition of two eigenstates} given by 
\beq 
\Psi(t<0) = \alpha_1\, \psi_1\, e^{-i E_1\,t/\hbar} +\alpha_2 \,\psi_2 \,e^{-i E_2\, t/\hbar}
\eeq{InitialP2}
where $|\alpha_1|^2 + |\alpha_2|^2=1$. After a delta-function pulse at $t=0$ one obtains the final state
\beq 
\Psi(t>0) = c_1(t>0)\, \psi_1\, e^{-i E_1\,t/\hbar} + c_2(t>0) \,\psi_2 \,e^{-i E_2\, t/\hbar}
\eeq{FinalP2}
where $c_1(t>0)$ and $c_2(t>0)$ are given by the expressions \reff{c1tg0} and \reff{c2tg0} respectively and are a function of $\alpha_1$, $\alpha_2$ and the interaction strength $\beta$. We will consider the transition from an initial state, a superposition of two eigenstates, to the single energy eigenstate $\psi_1$. We will determine what interaction strength $\beta$ leads to such a transition as a function of $\alpha_1$ and $\alpha_2$. 

A transition to the definite eigenstate $\psi_1$ means that in the final state one has $c_2(t>0)=0$ (and hence $|c_1(t>0)|^2=1$).  Using \reff{c2tg0}, this implies that  
\beq
\alpha _2 \cos \Big(\frac{| \beta | }{\hbar }\Big)-i\, \alpha_1 \,
\frac{|\beta|}{\beta } \, \sin \Big(\frac{| \beta | }{\hbar }\Big)=0
\eeq{Col}       
which yields the equation
\beq
\tan \Big(\frac{|\beta|}{\hbar}\Big)= \dfrac{-i \,\beta\,\alpha_2}{\alpha_1\,|\beta|}\,.
\eeq{Col2}
The left hand side of this equation is real which requires the right hand side to be real also. It is convenient to square the above equation by multiplying the right hand side by its complex conjugate. This yields
\begin{align} 
\tan^2 \Big(\frac{|\beta|}{\hbar}\Big)= \dfrac{|\alpha_2|^2}{|\alpha_1|^2}
=\dfrac{1-|\alpha_1|^2}{|\alpha_1|^2}=\dfrac{1}{|\alpha_1|^2}-1\,.
\end{align}
Since $\tan^2(x)+1=\sec^2(x)$ the above equation reduces to the simple equation
\beq
\cos \Big(\frac{|\beta|}{\hbar}\Big)=\pm |\alpha_1|\,.
\eeq{cos2}
It is convenient to define $k=\beta/\hbar$ where $k$ is a dimensionless quantity. So $k$ can be viewed as $\beta$ in units of $\hbar$. We will work with $k$ from now on instead of $\beta$ and refer to $k$ as the interaction strength. The solution to \reff{cos2} can be expressed as 
\beq
|k|=\cos^{-1}\big(\pm |\alpha_1|\big) \,.
\eeq{ModBeta}  
In $\cos^{-1}(x)$, $x$ can take on values ranging from $-1$ to $1$ inclusively. We define here $\cos^{-1}(x)$ to have a finite range from $0$ to $\pi$ inclusively ($0$ at $x=1$ and $\pi$ at $x=-1$). Since $\pm \,|\alpha_1|$ ranges from $-1$ to $1$, $\cos^{-1}(\pm\,|\alpha_1|)$ ranges between $0$ and $\pi$ which is non-negative. This is consistent with the fact that $|k|$ in \reff{ModBeta} is also non-negative. For now, $|k|$ ranges from $0$ to $\pi$ (we will extend this range later). For a given value of $|\alpha_1|$, there are two distinct values of $|k|$ in this range corresponding to the $\pm$ sign in \reff{ModBeta}: the positive sign yields a value of $|k|$ in the range between $0$ and $\pi/2$ inclusively and the negative sign yields a value of $|k|$ in the range between $\pi/2$ and $\,\pi$ inclusively ($|\alpha_1|=0$ is the exception and yields one value for $|k|$ which is $\pi/2$). 

Below we plot $|k|$ as a function of $|\alpha_1|$ in the range $0$ to $\pi$.   
\begin{figure}
	\centering
		\includegraphics[scale=0.90]{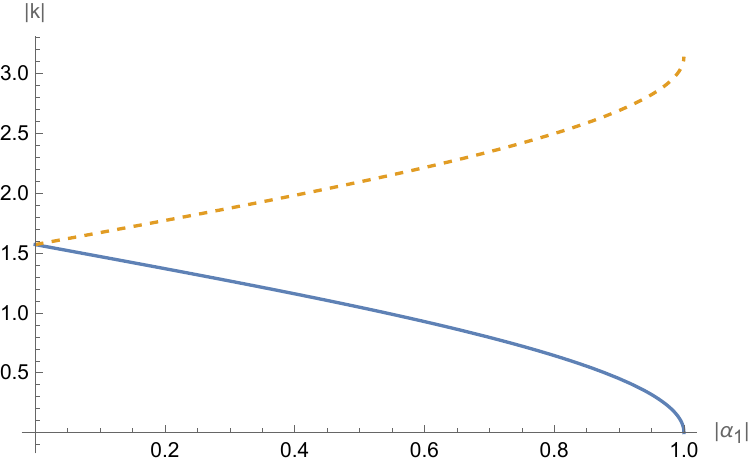}
		\caption{Plot of $|k|=|\beta|/\hbar$ vs. $|\alpha_1|$. The are two plots corresponding to the two signs in \reff{ModBeta}. The solid line plot corresponds to the positive sign and runs from $\pi/2$ to $0$ whereas the dashed plot corresponds to the negative sign and runs from $\pi/2$ to $\pi$.} 
	\label{fig:Betaplot}
\end{figure}
Note that for both plots, the slope becomes steeper as $\alpha_1$ increases towards unity. Let us quantify this by taking the derivative of \reff{ModBeta}:
\beq
\dfrac{d\,|k|}{d\,|\alpha_1|}=\mp \frac{1}{\sqrt{1-|\alpha_1|^2}}
\eeq{DerivK}
where the minus/plus sign corresponds to the solid/dashed plots respectively. For equal intervals $d|\alpha_1|$, the magnitude of $d|k|$ for both plots increases with $|\alpha_1|$ and reaches a maximum as $|\alpha_1|$ approaches $1$ (the slope diverges there). This implies that if one has a set of initial states whose modulus of $\alpha_1$ has a spread of $d|\alpha_1|$, the magnitude of $d|k|$ is greater if $|\alpha_1|$ is greater. There will therefore be a greater spread of possible values of the interaction that lead to a transition to the definite eigenstate $\psi_1$ if $|\alpha_1|$ is larger.  

We can extend the range of $|k|$ all the way to infinity by adding $2 \,\pi \,n$ to $\pm\, \cos^{-1}(\pm\,|\alpha_1|)$ where $n$ is a positive integer:  
\beq
 |k|=
\begin{cases}
&\cos^{-1}\big(\pm |\alpha_1|\big)\\
&2 \,\pi \,n \,\pm\, \cos^{-1}(\pm\,|\alpha_1|) \quad \forall \,n \in \mathbb{Z}^+ \,.
\end{cases} 
\eeq{betamag}    
The above gives all the possible values for the modulus of $k$ that leads to the definite eigenstate $\psi_1$. We have one more equation stemming from the fact that the right hand side of \reff{Col2} must be real. Replacing $\beta$ by $k\,\hbar$ this yields 
\beq
\dfrac{k}{k^*}=-\dfrac{\alpha_1}{\alpha_1^*}\,\dfrac{\alpha_2^*}{\alpha_2}\,.
\eeq{Reality}
For simplicity define the right hand side above as the complex number
\beq
z=-\dfrac{\alpha_1}{\alpha_1^*}\,\dfrac{\alpha_2^*}{\alpha_2}
\eeq{z}
and define $R$ as any of the values of $|k|$ given by the positive real quantities in \reff{betamag}. We therefore have $|k|=R$ and $k/k^*=z$. We can therefore solve for $k$. The final result for the interaction strength $k$ that yields a definite eigenstate is 
\beq
k= \pm R \,\sqrt{z}
\eeq{FinalBeta}
where $R$ has a value given by \reff{betamag} and $z$ is given by \reff{z}. The interaction strength $k$ is a function of $\alpha_1$ and $\alpha_2$ whereas its modulus $|k|$ is a function of $|\alpha_1|$ only. 

\subsection{A reversible process}

Since the abrupt/instantaneous transition to the definite eigenstate $\psi_1$ take place via the potential in Schr\"odinger's equation, the process should be reversible. That means we should be able to show that the same interaction (but negative) applied to the final state should bring us back to our original state. The final state is now the new initial state with coefficients labeled as $\alpha^{new}_1$ and $\alpha^{new}_2$. Since the final state corresponds to the first eigenstate only (no superposition) we have $\alpha^{new}_2=0$. We also have $|\alpha^{new}_1|=1$. We would however like to obtain $\alpha^{new}_1$ itself, not just its modulus. Recall that the initial superposition of states \reff{InitialP2} with coefficients $\alpha_1$ and $\alpha_2$ transitioned to the final state \reff{FinalP2} with coefficients $c_1(t>0)$ and 
$c_2(t>0)$ under the delta-function pulse. Since the final state was the first eigenstate this implied that $c_2(t>0)=0$. This condition led to the equations \reff{Col2} and \reff{cos2} for the $\beta$ that yields the final state. Though we know that 
$|c_1(t>0)|=1$ we now want to obtain $c_1(t>0)$ itself given by \reff{c1tg0} using the equations \reff{Col2} and \reff{cos2}. This will be $\alpha^{new}_1$:    
\begin{align}
c_1(t>0)&=\alpha _1 \cos \Big(\frac{| \beta | }{\hbar }\Big)-i \,\alpha_2 \,\frac{|\beta|}{\beta ^*}\, \sin \Big(\frac{| \beta | }{\hbar }\Big) \nonumber\\
&=\cos \Big(\frac{| \beta | }{\hbar }\Big)\,\Big(\alpha_1-i \,\alpha_2 \,\frac{|\beta|}{\beta ^*}\, \tan \Big(\frac{| \beta | }{\hbar }\Big)\, \Big)\nonumber\\
&=\pm |\alpha_1|\,\Big(\alpha_1 + \frac{|\alpha_2|^2}{\alpha_1^*} \Big)\nonumber\\
&=\pm \frac{\alpha_1}{|\alpha_1|}
\label{AlphaNew1}
\end{align}
where we used $|\alpha_2|^2=1-|\alpha_1|^2$. Note that $\beta$ above is the original interaction strength that led to the final state. The right hand side of \reff{Col2} is real and we used its complex conjugate instead in the above derivation.

Therefore the new initial state is given by
\beq 
\alpha^{new}_1=\pm \frac{\alpha_1}{|\alpha_1|} \text{ and } \alpha^{new}_2=0\,.
\eeq{AlphaNew}
After a new interaction $\beta^{new}$ acts on this new initial state we obtain the new final state with coefficients $c^{new}_1(t>0)$ and $c^{new}_2(t>0)$ given by \reff{c1tg0} and \reff{c2tg0} but with the new initial state coefficients $\alpha^{new}_1$ and $\alpha^{new}_2$ respectively. This yields 
\begin{align}
c^{new}_1(t>0)&=\alpha^{new}_1 \cos \Big(\frac{| \beta^{new} | }{\hbar }\Big)-i \,
\alpha^{new}_2 \,\frac{|\beta^{new}|}{\beta^{new^*}}\, \sin \Big(\frac{| \beta^{new} | }{\hbar }\Big)\nonumber\\
&=\pm \frac{\alpha_1}{|\alpha_1|}\,\cos \Big(\frac{| \beta^{new} | }{\hbar }\Big)
\label{cnew1}
\end{align}
and
\begin{align}
c^{new}_2(t>0)&=\alpha^{new}_2 \cos \Big(\frac{| \beta^{new} | }{\hbar }\Big)-i\, \alpha^{new}_1 \,
\frac{|\beta^{new}|}{\beta^{new}} \, \sin \Big(\frac{| \beta | }{\hbar }\Big)\nonumber\\
&=\mp\,i\,\frac{\alpha_1}{|\alpha_1|} \,
\frac{|\beta^{new}|}{\beta^{new}} \, \sin \Big(\frac{| \beta^{new} | }{\hbar }\Big)\,.
\label{cnew2}
\end{align}
Reversing the original process means that this new final state has to be equal to the original initial state: $c^{new}_1(t>0)=\alpha_1$ and $c^{new}_2(t>0)=\alpha_2$. We therefore obtain the following two equations:
\begin{align}
\pm \frac{\alpha_1}{|\alpha_1|}\,\cos \Big(\frac{| \beta^{new} | }{\hbar }\Big)&=\alpha_1\label{First}\\
\mp\,i\,\frac{\alpha_1}{|\alpha_1|} \,
\frac{|\beta^{new}|}{\beta^{new}} \, \sin \Big(\frac{| \beta^{new} | }{\hbar }\Big)&=\alpha_2\,.
\label{Second}     
\end{align}
The first equation \reff{First} yields 
\beq
\cos \Big(\frac{| \beta^{new} | }{\hbar }\Big)= \pm \,|\alpha_1|\,.
\eeq{cos1}
Dividing the second equation \reff{Second} by the first equation \reff{First} yields
\beq
\tan \Big(\frac{| \beta^{new} | }{\hbar }\Big)= \frac{i\,\beta^{new} \,\alpha_2}{\alpha_1\,| \beta^{new} |}\,.
\eeq{Col2B}
Comparing \reff{cos1} to \reff{cos2} and \reff{Col2B} to \reff{Col2} we see that
\beq
\beta^{new}=-\beta
\eeq{betanew}
where $\beta$ is the original interaction strength. By applying the negative of $\beta$ on the final state one recovers the initial state. This shows that the abrupt/instantaneous transition of a superposition of two eigenstates to a definite eigenstate using a delta-function pulse in Schr\"odinger's equation is a completely reversible process. In contrast, the collapse of the wavefunction induced by a measurement is an irreversible process even though it also involves an abrupt/instantaneous transition to a definite eigenstate.
 
\section{Conclusion}

In this paper we investigated general transitions under a delta-function pulse in a two-level system where the initial state was a linear superposition of two eigenstates. This system was solved exactly using two different methods (one of them presented in appendix A) and the general analytical expressions \reff{c1tg0} and \reff{c2tg0} were obtained for the coefficients $c_1(t>0)$ and $c_2(t>0)$ respectively in the final state. These expressions were a function of the initial coefficients $\alpha_1$ and $\alpha_2$ as well as the interaction strength $\beta$. They are general in the sense that one could choose any value for $\alpha_1$ and $\alpha_2$ (within the constraint that $|\alpha_1|^2+|\alpha_2|^2=1$). Previous work had focused mainly on the transition probability $P_{1\to 2}$ in going from the first to the second eigenstate. We obtained a general expression $P_{\alpha_1,\alpha_2 \to 2}$ given by \reff{c2Sq} for the transition probability to the second eigenstate starting from a general initial state. This opened up new possibilities where under a certain interaction strength, the sign of the initial state flips or the roles played by $|\alpha_1|^2$ and $|\alpha_2|^2$ in the initial state are switched in the final state (e.g. $|\alpha_1|^2$ becomes probability of measuring $E_2$ instead of $E_1$).     

Under a delta-function pulse, we were able to investigate a scenario where an initial superposition of two eigenstates transitions abruptly to a definite eigenstate i.e. with unit probability. Without loss of generality, we studied the case where the definite eigenstate was the first eigenstate. Using the analytical expressions for the coefficients $c_1(t>0)$ and $c_2(t>0)$ we showed that starting with a general initial state the values of the interaction strength $\beta$ that lead to the first eigenstate obey the equations \reff{Col2} and \reff{cos2} with the modulus $|k|=|\beta|/\hbar$ depending on $|\alpha_1|$ only. We plotted $|k|$ as a function of $|\alpha_1|$ where $|\alpha_1|$ runs from $0$ to $1$ inclusively. The magnitude of the slope in the plot increases with $|\alpha_1|$ which means that for a spread $d|\alpha_1|$ of a set of initial states, the spread $d|k|$ in interaction strengths that lead to the first eigenstate (the definite eigenstate) is greater when $|\alpha_1|$ is larger. Simply put, there is a greater spread of possible values of the interaction that lead to the first eigenstate in the initial set if $|\alpha_1|$ is larger.

The abrupt transition from an initial superposition of eigenstates to a definite eigenstate under a delta-function pulse is of particular interest because it shares two aspects in common with the collapse of the wavefunction induced by a measurement. First, the wavefunction is also a linear superposition of eigenstates and secondly, it also transitions abruptly or instantaneously (collapses) to a definite eigenstate. However, an important difference is that you do not know a priori which eigenstate you will end up in after a measurement. All you know is that the probability of ending up in a given eigenstate is governed by the Born rule. A measurement is also an irreversible process whereas we showed explicitly in section 4 that the action of a delta-function pulse is a reversible process.       

An important and interesting feature of a delta-function pulse is that it leads to transition probabilities that do not depend on the energy gap $\Delta E=E_2-E_1$. There is a nice physically intuitive way to view this independence. The Fourier transform $\tilde{f}(\omega)$ of a Dirac delta-function $f(t)=\delta (t)$ is constant and equal to unity meaning it covers \textit{all} frequencies $\omega=\Delta E/\hbar$ with \textit{equal} weight. This implies the probability is flat with respect to $\Delta E$. In section 3 we saw explicitly that the dependence on the energy gap that exists for finite pulse widths disappears in the limit as the width tends to zero. This implies that the relative phase $e^{i \,\omega_0 \,t}$ where $\omega_0=(E_2-E_1)/\hbar$ has no effect under a delta-function pulse. This brings to mind the decoherence viewpoint (for a review see \cite{Schloss}), where the interaction of a pure state with the environment leads to a rapid, exponentially decreasing loss of the phase relationships in the pure state. The loss is thought to occur rapidly due to the large number of degrees of freedom involved in the macroscopic environment and measuring apparatus. The delta-function pulse and decoherence therefore have the loss of relative phase in common that stems from a short intense interaction. The similarities however stop here. Under a delta-function pulse the system does not lose information and the system undergoes changes that are reversible while in decoherence information is lost to the environment and the system undergoes changes that are irreversible. This is why decoherence can offer an explanation as to why a measurement is an irreversible process. After decoherence, the system behaves as a classical statistical ensemble rather than as a single coherent quantum superposition. This is a major step forward because what we actually measure is an eigenvalue associated with the projection to a single eigenstate and this can be viewed as a classical outcome. However, decoherence has yet to describe adequately the mechanism that leads a measurement to select a \textit{single} outcome out of many and with a probability governed by the Born rule. 


\section*{Data Avalilability} 

No data was used for the research described in the article.

\section*{Acknowledgments}
The author thanks Bishop's University for their financial support.  

\begin{appendices}
\numberwithin{equation}{section}
\setcounter{equation}{0}
\section{A different method for solving the coupled set of first order equations}

In section \ref{S3} we solved the coupled set of first order equations \reff{c1dot2} and \reff{c2dot2} that mix the coefficients $c_1(t)$ and $c_2(t)$ by rewriting them as a second order equation \reff{c1Second2} for $c_1(t)$. In the large (infinite) $n$ limit (delta-function pulse) we obtained the final expressions \reff{c1tg0} and \reff{c2tg0} for $c_1(t>0)$ and $c_2(t>0)$ respectively. In this section we solve the coupled set of first order equations using a different method. Instead of combining them into a second order equation we solve two separate first order equations. We obtain the same expressions as \reff{c1tg0} and \reff{c2tg0} for $c_1(t>0)$ and $c_2(t>0)$ respectively. For ease of reference we rewrite the coupled equations here:
\beq
\dot{c_1}=\frac{-i}{\hbar}\Big(c_2 \,\beta\, q_n(t)\, e^{-i\,\omega_0\, t}\Big)
\eeq{c1dotA}
\beq 
\dot{c_2}=\frac{-i}{\hbar}\Big(c_1 \,\beta^*\, q_n(t)\, e^{i\,\omega_0 \,t}\Big)\,.
\eeq{c2dotA}  

In the large $n$ limit, $q_n(t)$ is concentrated near $t=0$. The functions $e^{\pm\, i \,\omega_0\,t}$ appearing in \reff{c1dotA} and \reff{c2dotA} are smooth functions that are perfectly well-behaved at $t=0$ so that in the large $n$ limit, they can be set to unity. The error in this approximation decreases as $n$ gets larger and vanishes for the delta-function pulse we consider (corresponding to the infinite $n$ limit). We begin by dividing \reff{c1dotA} by \reff{c2dotA} and obtain    
\begin{align}
\dfrac{\dot{c_1}}{\dot{c_2}}&= \dfrac{c_2 \,\beta}{c_1 \,\beta^*}\nonumber \\
\beta^*\,\dot{c_1}\,c_1 &=\beta\,\dot{c_2}\,c_2 \,\nonumber\\
\beta^*\,\dfrac{d\,c_1^2}{dt}&=\beta\,\dfrac{d\,c_2^2}{dt}\,.\label{dc2}
\end{align}
Integrating both sides of \reff{dc2} yields
\begin{align}
\beta^*\,\int_{-\infty}^t\dfrac{d\,c_1^2}{dt'}\,dt'\,&=\,\beta\,\int_{-\infty}^t\dfrac{d\,c_2^2}{dt'}\,dt'
\nonumber\\
\beta^*\,\int_{-\infty}^t d\,c_1^2&=\beta\,\int_{-\infty}^t d\,c_2^2\nonumber\\
\beta^*\,(\,c_1(t)^2-\alpha_1^2\,)&= \beta\,(\,c_2(t)^2 -\alpha_2^2\,)
\label{CSquare}
\end{align}
where we used $c_1(-\infty)=\alpha_1$ and $c_2(-\infty)=\alpha_2$. Using 
\reff{CSquare} we can express $c_2(t)$ in terms of $c_1(t)$:
\beq
c_2(t)= \pm\,\sqrt{\,(\,c_1(t)^2-\alpha_1^2\,)\,\tfrac{\beta^*}{\beta} + \alpha_2^2}\,.
\eeq{c2Root}
It is cumbersome to carry the $\pm$ sign above in the calculations so we will simply consider the positive sign below. It can be checked that the negative sign yields the same results when the initial conditions are satisfied. Substituting $c_2$ given by \reff{c2Root} and $q_n(t)$ given by the Gaussians 
\reff{func} into \reff{c1dotA} one obtains the differential equation
\beq
\dot{c_1}=\frac{- i\,\beta}{\hbar}\,\sqrt{\,(\,c_1(t)^2-\alpha_1^2\,)\,\tfrac{\beta^*}{\beta} + \alpha_2^2}\,\,\frac{n}{\sqrt{\pi}}\, e^{-n^2 \,t^2} \, e^{-i\,\omega_0\, t}\,.
\eeq{c1Diff}
We have kept the $e^{-i\,\omega_0\, t}$ term here (instead of setting it to unity) in order to illustrate that even when we include it at this point, $\omega_0$ will not enter into the result in the large (infinite) $n$ limit. The solution to \reff{c1Diff} is
\begin{align}
c_1(t)&=\frac{1}{2} \left(\alpha _1^2 \beta ^*-\alpha _2^2 \beta \right) \exp \Big( \,\frac{i\,|\beta| \, e^{-\frac{\omega _0^2}{4 n^2}} \text{erf}\left(n t+\frac{i \omega _0}{2 n}\right)}{2\,\hbar }- \,d\,\sqrt{\beta ^*}\Big)\nonumber\\&\quad+\frac{1}{2 \beta ^*}\exp \Big(-\frac{i\,|\beta | e^{-\frac{\omega _0^2}{4 n^2}} \text{erf}\left(n t+\frac{i \omega _0}{2 n}\right) }{2\,\hbar }+d\,\sqrt{\beta ^*}\Big)
\label{c1sol}
\end{align}
where $d$ is an integration constant. In the large (infinite) $n$ limit,  the terms involving $\omega_0$ can be removed as expected and the above expression reduces to   
\begin{align}
c_1(t)&=\frac{1}{2}\,( \alpha _1^2\, \beta ^* -\alpha _2^2 \,\beta)\, \exp \left(-d\, \sqrt{\beta ^*}+\,\frac{i \,| \beta |}{2\, \hbar}  \,\text{erf}\left(n \,t\right)\right)\nonumber\\
&+\frac{1}{2 \,\beta ^*}\,\exp \left(d\, \sqrt{\beta ^*}-\,\frac{i \,| \beta |}{2\, \hbar}\, \, \text{erf}\left(n \,t\right)\right)\,.
\label{c1sol2}
\end{align}
Substituting $c_1(t)$ above into the original equation \reff{c1dotA} we obtain an expression for $c_2(t)$:
\begin{align}
c_2(t)\!=\!\frac{1}{2 \,| \beta | }\bigg(\!\beta ^* \left(\alpha _2^2 \beta -\alpha _1^2 \beta ^*\right) \exp \Big(\!-d\, \sqrt{\beta ^*}+\frac{i \,| \beta |}{2\, \hbar} \,\text{erf}\left(n \,t\right)\!\Big)\!+\!\exp \Big(\!d\, \sqrt{\beta ^*}-\frac{i \,| \beta |}{2\, \hbar}\, \, \text{erf}\left(n \,t\right)\!\Big)\!\bigg)
\label{c2sol2}
\end{align}
where again we considered the large $n$ limit where $e^{-i\,\omega_0\, t}$ can be set to unity. We can determine $d$ by requiring that $c_1(t<0)=\alpha_1$ and $c_2(t<0)=\alpha_2$. When $t<0$, $\lim_{n\to \infty} \text{erf}(n\,t)=-1$. Substituting this back into \reff{c1sol2} leads to the following equation for $c_1(t<0)$: 
\begin{align} 
c_1(t<0)&=\frac{1}{2}\,( \alpha _1^2\, \beta ^* -\alpha _2^2 \,\beta)\, \exp \left(-d\, \sqrt{\beta ^*}- \frac{i \,| \beta |}{2\, \hbar} \right)
+\frac{1}{2 \,\beta ^*}\,\exp \left(d\, \sqrt{\beta ^*}+\frac{i \,| \beta |}{2\, \hbar}\, \, \right)\nonumber\\
&=\alpha_1\,.
\label{c1alpha1}
\end{align}
The above equation has two solutions for $d$:  
\begin{align}
d&=\frac{2\,\hbar \, \big(2\, i \,\pi \, k +\log \left(\,\beta ^* \alpha _1-| \beta |\,\alpha _2\,\right)\big)-i \,| \beta | }{2 \,\hbar  \,\sqrt{\beta ^*}}\text{ where }k \in \mathbb{Z}\label{d1}\\
d&= \frac{2 \,\hbar \, \big(2\, i\, \pi \,k+\log \left(\,\beta ^* \alpha _1+| \beta |\, \alpha _2\,\right)\big)-i\, | \beta | }{2\, \hbar \, \sqrt{\beta ^*}}\text{ where }k \in \mathbb{Z}\label{d2}\,.
\end{align}
It can be checked that the solution \reff{d2} for $d$ satisfies also the condition $c_2(t<0)=\alpha_2$ (the other solution \reff{d1} yields $c_2(t<0)=-\alpha_2$ which is not the desired initial condition). 

Our goal is to obtain $c_1(t)$ and $c_2(t)$ for $t>0$, that is after the delta-function pulse at $t=0$ has acted. For $t>0$, $\lim_{n\to \infty} \text{erf}(n\,t)=1$. Substituting that result as well as the solution \reff{d2} for the constant $d$ into \reff{c1sol2} and \reff{c2sol2} yields respectively
\beq
c_1(t>0)=\alpha _1 \cos \left(\frac{| \beta | }{\hbar }\right)- i\, \alpha _2 \sqrt{\tfrac{\beta}{\beta ^*}} \sin \left(\frac{| \beta | }{\hbar }\right)
\eeq{c1t0}
and
\beq
c_2(t>0)=\alpha _2 \cos \left(\frac{| \beta | }{\hbar }\right)- i\, \alpha _1 \sqrt{\tfrac{\beta^*}{\beta}} \sin \left(\frac{| \beta | }{\hbar }\right)\,.
\eeq{c2t0}
The above exact expressions for $c_1(t>0)$ and $c_2(t>0)$ are the same as the expressions \reff{c1tg0} and \reff{c2tg0} respectively that we obtained in section \ref{S3}. This provides a strong confirmation for our exact expressions. 
\end{appendices}

\end{document}